# Secure and *Practical* Identity-Based Encryption

David Naccache

Groupe de Cyptographie, Département d'Informatique
École Normale Supérieure
45 rue d'Ulm, 75005 Paris, France
`david.nacache@ens.fr`

**Abstract.** In this paper, we present a variant of Waters' Identity-Based Encryption scheme with a much smaller public-key size (only a few kilobytes). We show that this variant is semantically secure against passive adversaries in the standard model.

In essence, the new scheme divides Waters' public key size by a factor $\ell$ at the cost of (negligibly) reducing security by $\ell$ bits. Therefore, our construction settles an open question asked by Waters and constitutes the first fully secure *practical* Identity-Based Encryption scheme[1].

## 1 Introduction

The concept of Identity-Based Encryption (IBE) was invented by Shamir in 1984 [9]. It allows a party to encrypt a message using the recipient's identity as a public key. The corresponding private-key is provided by a central authority. The advantage of IBE over conventional public-key encryption is that it avoids certificate management, which greatly simplifies the implementation of secure communications between users. With an IBE scheme, users can simply use their email addresses as their identities. Moreover, the recipient does not need to be online to present a public-key certificate before the sender encrypts a message, and the sender does not have to be online to check the validity of the certificate.

There are currently two IBE security notions. The stronger notion, called *semantic security against passive adversaries* (IND-ID-CPA), was introduced by Boneh and Franklin in [2]. As per this notion, the adversary can request the private keys for identities of his choosing; eventually he must be able to distinguish the encryption of two messages for an identity he decides of[2]. This notion will be described in detail in section 2. A weaker notion of security, introduced by Canetti, Halevi and Katz in [5,6], is called *selective-ID semantic security against passive adversaries* (IND-sID-CPA). As per this notion, the adversary must commit ahead of time to the identity that it will attack, that is, *before* he receives the public parameters. In this paper, an IBE scheme satisfying the stronger notion will be called *fully secure*. The present paper's goal is to construct a practical and fully secure IBE scheme.

---

[1] Work done while the author was employed by Gemplus (Gemplus patent pending)
[2] Of course, different from the identities whose private keys he requested.

There are several security models for public-key cryptosystems. The *random oracle model* has been introduced by Bellare and Rogaway as a "paradigm for designing efficient protocols" [1]. It assumes that all parties, including the adversary, have access to a public, truly random hash function $H$. In practice, this ideal hash function $H$ is instantiated as a concrete cryptographic hash function (for example, SHA-1 [8]). This model proved to be extremely useful for designing simple, efficient and highly practical solutions for many problems. However, from a theoretical perspective, it is clear that a security proof in the random oracle model is only a heuristic indication of the system's security when instantiated with a particular hash function such as SHA-1.

*A contrario*, in the *standard model*, one does not assume idealized oracle accesses. In the standard model, security is proven using only standard complexity assumptions. Consequently, from a security perspective, a proof in the standard model is preferable to a proof in the random oracle model. Therefore, an important research direction in modern cryptography is the construction of cryptosystems provably secure in the standard model, with an efficiency comparable to what can be achieved in the random oracle model. In this article an IBE scheme will be (*sub silentio*) called *fully secure* if it is fully secure in the standard model. Our goal is to construct a practical and fully secure IBE.

The first efficient identity-based encryption scheme was proposed by Boneh and Franklin at CRYPTO 2001 [2]. The Boneh-Franklin scheme is fully secure in the random oracle model. This scheme and all subsequent IBE schemes (except one exotic species [7]) are based on bilinear maps; the only known construction of bilinear maps is based on the Weil or the Tate pairing over certain families of elliptic curves. Since then, Boneh and Boyen proposed at EUROCRYPT 2004 an IBE scheme secure without random oracles, but reaching only selective ID security [3]. Boneh and Boyen also proposed at CRYPTO 2004 an IBE scheme fully secure in the standard model (*i.e.* without random oracles), but their scheme is too inefficient to be practical. Finally, the first practical and fully secure IBE scheme was proposed at EUROCRYPT 2005 by Waters [10]. Encryption and decryption are very efficient since only a few exponentiations and bilinear map computations are required.

However, a drawback of Waters' scheme is that the size of the public parameters is very large: namely, the public parameters contain $n+4$ group elements, where $n$ is the size of the bit-string representing identities. Since $n$ can be the output of a hash-function, we must take at least $n = 160$. Moreover, when an Elliptic Curve over $\mathbb{F}_p^2$ is used (the simpler setting), the size of a group element must be at least 1024 bits, to attain a security level equivalent to a 1024-bit RSA. Therefore, each participant must store at least 164 kilobytes of public parameters, which is prohibitive for most present-day "normal" smart card. In the conclusion of his EUROCRYPT 2005 paper [10], Waters states that finding an efficient identity-based encryption scheme secure without random oracles with short public parameters is an open problem.

This paper solves this open problem and introduces a variant of Waters' scheme with a much smaller public-key size (only a few kilobytes). *Eo ipso*, we define the first *practical identity based encryption Scheme, semantically secure against passive adversaries in the standard model*.

## 2 Definitions

In this section we first recall the definition of an IBE scheme. We then recall the definition of semantic security against passive adversaries for IBE, introduced in [2].

An IBE consists scheme of four algorithms :

- Setup : the Setup algorithm generates the system's public parameters, denoted by *params*, and a private master key denoted *master-key*.
- Keygen : the Keygen algorithm takes as input an identity $v$ and outputs the private key $d_v$ for identity $v$, using *master-key*.
- Encrypt : the encryption algorithm encrypts messages for an identity $v$ using *params*.
- Decrypt : the decryption algorithm decrypts ciphertexts for identity $v$ using the private-key $d_v$.

The semantic security of an IBE scheme is defined through the following scenario between an attacker $\mathcal{A}$ and a challenger $\mathcal{C}$ :

- *Setup* : $\mathcal{C}$ generates the master public parameters and gives them to $\mathcal{A}$.
- *Phase 1* : $\mathcal{A}$ can request the private-key corresponding to an identity $v$ of his choice. $\mathcal{A}$ can repeat this multiple times for different identities.
- *Challenge* : $\mathcal{A}$ submits an identity $v^*$, different from the identities in Phase 1, and two messages $m_0$ and $m_1$. $\mathcal{C}$ flips a coin $b$ and returns the encryption of $m_b$ under identity $v^*$.
- *Phase 2* : Phase 1 is repeated with the restriction that $\mathcal{A}$ cannot request the private key for $v^*$.
- *Guess* : $\mathcal{A}$ submits a guess $b'$ for $b$.

This completes the description of the scenario. The advantage of an adversary $\mathcal{A}$ in breaking the scheme is defined as :

$$\text{Adv}(\mathcal{A}) = \left| \Pr[b' = b] - \frac{1}{2} \right|$$

where the probability is taken over the adversary's random coins and the challenger's random coins.

**Definition 1 (IBE semantic security).** *An IBE scheme is said to be $(t, q, \varepsilon)$-semantically secure if all $t$-time adversaries making at most $q$ private key queries have an advantage at most $\varepsilon$ in breaking the scheme.*

## 3 Complexity Assumptions

The new construction is based on bilinear maps. In this section, we recall known facts and complexity assumptions on bilinear maps; the reader is referred to [2] for more details.

Let $\mathbb{G}$ and $\mathbb{G}_1$ be groups of prime order $p$ and let $g$ be a generator of $\mathbb{G}$. We say that $\mathbb{G}$ has a bilinear map $e : \mathbb{G} \times \mathbb{G} \to \mathbb{G}_1$ if the following conditions hold : $e$ is efficiently computable, $e$ is bilinear, that is $e(g^a, g^b) = e(g,g)^{ab}$ for all $a, b$ and $e$ is non-degenerate, that is $e(g,g) \neq 1$.

The Bilinear Diffie-Hellman problem is defined as follows :

**Definition 2 (Bilinear Diffie-Hellman Problem (BDH)).** *Given the 4-uple $(g, g^a, g^b, g^c)$ where $a, b, c \leftarrow \mathbb{Z}_p$, output $e(g,g)^{abc}$.*

The decisional version is defined in the usual manner :

**Definition 3 (Decisional Bilinear Diffie-Hellman Problem (DBDH)).** *Let $g, g^a, g^b, g^c$ defined as previously. Let $\beta$ be a random binary coin. Let $z = e(g,g)^{abc}$ if $\beta = 1$, and let $z$ be a random element in $\mathbb{G}_1$ otherwise. Given $(g, g^a, g^b, g^c, z)$, output a guess $\beta'$ of $\beta$.*

We say that an algorithm as an advantage $\varepsilon$ in solving DBDH if

$$\left| \Pr[\beta' = \beta] - \frac{1}{2} \right| \geq \varepsilon$$

**Definition 4 (DBDH Assumption).** *We say that the $(t, \varepsilon)$-DBDH assumption holds in $\mathbb{G}$ if no $t$-time algorithm has an advantage at least $\varepsilon$ in solving the DBDH problem in $\mathbb{G}$.*

## 4 The New Idea

The following describes a new practical and fully secure IBE scheme.

The new scheme is a variant of Waters' IBE, but with shorter public parameters. Let $\mathbb{G}$ be a group of prime order $p$, let $g$ be a generator of $\mathbb{G}$, and let $e$ be an admissible bilinear map into $\mathbb{G}_1$. Identities will be represented as $n$ dimensional vectors $v = (v_1, \ldots, v_n)$ where each $v_i$ is an $\ell$-bit integer. The integers $n$ and $\ell$ are parameters unrelated to $p$, and $n' = n \cdot \ell$ is the output length of a collision-resistant hash function $H : \{0,1\}^* \to \{0,1\}^{n'}$.

Setup : A secret $\alpha \in \mathbb{Z}_p$ is chosen at random. One sets $g_1 = g^\alpha$ and $g_2$ is chosen randomly in $\mathbb{G}$. One chooses a random $u' \in \mathbb{G}$ and a random $n$ dimensional vector $U = (u_i)$ whose elements are randomly chosen in $\mathbb{G}$. The public parameters are $g, g_1, g_2, u'$ and $U$. The master secret is $g_2^\alpha$.

Keygen : Let $v = (v_1, \ldots, v_n) \in (\{0,1\}^a)^n$ be an identity. Let $r$ be random in $\mathbb{Z}_p$. The private key $d_v$ for identity $v$ is constructed as :

$$d_v = \left( g_2^\alpha \left( u' \prod_{i=1}^n u_i^{v_i} \right)^r, g^r \right) \quad (1)$$

Encryption : A message $m$ is encrypted for identity $v$ as follows. A value $t$ is chosen at random in $\mathbb{Z}_p$. The ciphertext is then :

$$c = \left( e(g_1, g_2)^t \cdot m, g^t, \left( u' \prod_{i=1}^n u_i^{v_i} \right)^t \right)$$

Decryption : Let $c = (c_1, c_2, c_3)$ be an encryption of $m$ under identity $v$. The ciphertext $c$ can then be decrypted using $d_v = (d_1, d_2)$ as :

$$\begin{aligned}
c_1 \frac{e(d_2, c_3)}{e(c_2, d_1)} &= e(g_1, g_2)^t \cdot m \cdot \frac{e(g^r, (u' \prod u_i^{v_i})^t)}{e(g^t, g_2^\alpha (u' \prod u_i^{v_i})^r)} \\
&= e(g_1, g_2)^t \cdot m \cdot \frac{e(g, (u' \prod u_i^{v_i})^{rt})}{e(g_1, g_2)^t e(g, (u' \prod u_i^{v_i})^{rt})} \\
&= m
\end{aligned}$$

### 4.1 How Does The New Cryptosystem Relate to Waters' Scheme?

Our construction is a modification of the Waters' scheme [10]. Namely, in Waters' scheme, to encrypt a message for identity $v = (v_1, \ldots, v_{n'}) \in \{0,1\}^{n'}$, one computes the product :

$$u' \cdot \prod_{v_i=1} u_i$$

where $U = (u_1, \ldots, u_{n'})$ is an $n'$ dimensional public vector.

The new construction encodes identities as a $n$ dimensional vectors $v = (v_1, \ldots, v_n)$ where each $v_i$ is a $\ell$-bit integers and $n \cdot \ell = n'$, and computes the modified product :

$$u' \cdot \prod_{i=1}^n u_i^{v_i}$$

where $U = (u_1, \ldots, u_n)$ is now an $n$ dimensional public vector. Therefore, the size of the public vector $U$ is slashed by a factor $n'/n = \ell$.

### 4.2 Performance

The size of the public parameter is $n + 4$ group elements, where $n' = n \cdot \ell$ is the output size of a collision-resistant hash-function. If the value $e(g_1, g_2)$ is precomputed, encryption requires the equivalent of one exponentiation in $\mathbb{G}_1$ and three exponentiations in $\mathbb{G}$. Decryption requires two bilinear map computations, one group operation $\mathbb{G}_1$ and one inversion in $\mathbb{G}_1$.

Compared to Waters' scheme, the public parameter size is shrunk by a factor $\ell$; encryption and decryption are almost as efficient as in Waters' scheme.

## 5  Security Proof

The following theorem proves that the new cryptosystem is fully secure in the standard model, under the Decisional Bilinear Diffie-Hellman Assumption.

**Theorem 1.** *The new* IBE *construction is* $(t, q, \varepsilon)$-*semantically secure, assuming that the* $(t', \varepsilon')$-DBDH *assumption holds, where :*

$$t = t' - \mathcal{O}(\varepsilon^{-2} \ln(\varepsilon^{-1}) \lambda^{-1} \ln(\lambda^{-1}))$$
$$\varepsilon = q \cdot 2^{\ell+4} \cdot n \cdot \varepsilon'$$

where $\lambda = 1/(q \cdot 2^{\ell+2} \cdot n)$.

*Proof.* The security proof is very similar to Waters' proof given in [10].

Assume that there exists a $(t, q, \varepsilon)$ adversary $\mathcal{A}$. We construct a simulator $\mathcal{B}$ that solves the DBDH problem with advantage at least $\varepsilon'$. The simulator $\mathcal{B}$ receives the DBDH challenge $(g, A = g^a, B = g^b, C = g^c, z)$ and must output a guess $\beta'$ as to whether $z = e(g,g)^{abc}$ ($\beta = 1$) or $z$ is a random element in $\mathbb{G}_1$ ($\beta = 0$). As in [10] we first describe a simulator $\mathcal{B}$ that does not quite work, and then we slightly modify it so that it works.

*Setup*: The simulator $\mathcal{B}$ first sets an integer $m = 2q$ and chooses randomly:

| | |
|---|---|
| an integer $k$ | $(0 \leq k \leq 2^\ell \cdot n - 1)$. |
| an $n$ dimensional vector $x = (x_1, \ldots, x_n)$ | (where $0 \leq x_i \leq m - 1$) |
| an integer $x'$ | (where $0 \leq x' \leq m - 1$). |
| a $y \in \mathbb{Z}_p$ | |
| and an $n$ dimensional vector $y = (y_1, \ldots, y_n)$ | (where each $y_i \in \mathbb{Z}_p$) |

For an identity $v = (v_1, \ldots, v_n)$, we define the functions :

$$F(v) = x' + \sum_{i=1}^{n} v_i \cdot x_i - m \cdot k$$
$$J(v) = y' + \sum_{i=1}^{n} v_i \cdot y_i$$

The simulator lets $g_1 = A$ and $g_2 = B$. It then defines the public parameters $u' = g_2^{x'-km} g^{y'}$ and $u_i = g_2^{x_i} g^{y_i}$. Therefore, the distribution of the public parameters is the same as in the attack scenario.

We have that for any identity $v$ :

$$u' \cdot \prod_{i=1}^{n} u_i^{v_i} = g_2^{F(v)} g^{J(v)} \qquad (2)$$

The following equality, valid if $F(v) \neq 0 \mod p$, will be used to answer the private-key queries :

$$\left(u' \cdot \prod_{i=1}^{n} u_i^{v_i}\right)^{\frac{a}{F(v)}} = g_2^a \cdot g_1^{J(v)/F(v)} \qquad (3)$$

*Phase 1*: The simulator $\mathcal{B}$ must answer the private key queries of $\mathcal{A}$. Consider a query for identity $v$. To answer this query, it would be sufficient to have $F(v) \neq 0 \mod p$. We observe that since $-p < F(v) < p$, we have that $F(v) = 0 \mod p$ implies $F(v) = 0$ and therefore $F(v) = 0 \mod m$. Here we only answer the private-key query if $F(v) \neq 0 \mod m$, which implies $F(v) \neq 0 \mod p$. In this case, $\mathcal{B}$ generates a random $r$ in $\mathbb{Z}_p$ and constructs the private key $d_v$ as follows :

$$d_v = \left(g_1^{-J(v)/F(v)} (u' \cdot \prod_{i=1}^{n} u_i^{v_i})^r, g_1^{-1/F(v)} g^r\right)$$

Letting $r' = r - a/F(v)$ and using (3), one obtains :

$$d_v = \left(g_2^a (u' \cdot \prod_{i=1}^{n} u_i^{v_i})^{r'}, g^{r'}\right)$$

which shows that $d_v$ is a valid private key for identity $v$, with the same distribution as in the attack scenario.

Otherwise, if $F(v) = 0 \mod m$, then the simulator $\mathcal{B}$ aborts and outputs a random bit $\beta'$ as its guess for $\beta$.

*Challenge*: The adversary submits two message $m_0$ and $m_1$ and an identity $v^*$. Again, we distinguish two cases : if $F(v^*) \neq 0 \mod p$, then the simulator $\mathcal{B}$ aborts and outputs a random bit $\beta'$. Otherwise, the simulator flips a fair binary coin $\gamma$ and constructs the ciphertext :

$$T = (z \cdot m_\gamma, C, C^{J(v^*)})$$

If $\mathcal{B}$ was given a legitimate BDH tuple, *i.e.* if $z = e(g,g)^{abc}$, then by virtue of equation (2) and $F(v^*) = 0 \mod p$ we have :

$$T = \left(e(g,g)^{abc} \cdot m_\gamma, g^c, g^{cJ(v^*)}\right) = \left(e(g_1,g_2)^c \cdot m_\gamma, g^c, (u' \prod u_i^{v_i})^c\right)$$

This shows that the ciphertext $T$ is a valid encryption of $m_\gamma$ with the same distribution as in the attack scenario. Otherwise, we have that $z$ is a random element of $\mathbb{G}_1$, which implies that the adversary obtains no information about $\gamma$ whatsoever.

*Phase 2*: The simulator $\mathcal{B}$ repeats the same operation as in Phase 1.

*Guess*: The adversary outputs a guess $\gamma'$ for $\gamma$. If $\gamma = \gamma'$, the simulator $\mathcal{B}$ outputs $\beta' = 1$, otherwise it outputs $\beta' = 0$.

This completes the description of simulator $\mathcal{B}$. As in [10], the problem with simulator $\mathcal{B}$ is that it aborts with a probability that is a function of the queried identities $v$ and $v^*$. Therefore, even if $\Pr[\gamma' = \gamma] > \frac{1}{2} + \varepsilon$ in the real attack scenario, we might have $\Pr[\beta' = \beta] \simeq \frac{1}{2}$ in the simulation. As in [10], a solution consists in artificially aborting the simulator at the end of the guess phase, so that the overall probability of aborting is made nearly constant.

First, we analyse the probability that the simulator $\mathcal{B}$ aborts when answering a private-key query or in the challenge phase. To start with, we fix all the random variables that the adversary can see, including its random coins: we fix the BDH tuple $(g, A = g^a, B = g^b, C = g^c, z)$, the public parameters $u'$ and the $u_i$. We also fix the random numbers $r'$ used when answering private-key queries, and the binary coin $\gamma$. Let $\mathfrak{P}$ denote those fixed parameters. The adversary can then be seen as a deterministic algorithm. In particular, the identities $v^j$, $1 \leq j \leq q$ queried by the adversary are fixed, and also the challenge identity $v^*$.

We observe that when those random variables are fixed, the random variables $x'$ and $x_i$ still have an independent and uniform distribution between zero and $m-1$, and $k$ has a uniform distribution between zero and $2^\ell \cdot n - 1$. We let $\boldsymbol{V} = (v^1, \ldots, v^q)$ be the list of private-key queries and let $\boldsymbol{X} = (x', x_1, \ldots, x_n)$.

We define the function :
$$\tau(\boldsymbol{X}, \boldsymbol{v}, v^*, k) = \begin{cases} 0, & \text{if } F(v^*) = 0 \text{ and } F(v^j) \neq 0 \mod m \text{ for all } 1 \leq j \leq q \\ 1, & \text{otherwise} \end{cases}$$

We have that the simulator $\mathcal{B}$ does not abort iff $\tau(\boldsymbol{X}, \boldsymbol{v}, v^*, k) = 0$. In Appendix A, we show the following lower bound for the probability that $\mathcal{B}$ does not abort :
$$\Pr_{\boldsymbol{X}, k}[\tau(\boldsymbol{X}, \boldsymbol{v}, v^*, k) = 0] \geq \lambda = \frac{1}{4 \cdot q \cdot 2^\ell \cdot n}$$

In the following, we modify the simulator so that it always aborts with probability nearly $\lambda$. In the guess phase, the new simulator $\mathcal{B}'$ will sample an estimate $\eta'$ of the probability $\Pr_{\boldsymbol{X}, k}[\tau(\boldsymbol{X}, \boldsymbol{v}, v^*, k) = 0]$. This probability is a function of $\boldsymbol{v}$ and $v^*$. Then if $\eta' > \lambda$, simulator $\mathcal{B}'$ will proceed as in $\mathcal{B}$ with probability $\lambda/\eta'$, and artificially abort with probability $1 - \lambda/\eta'$; in this latter case, it outputs a random guess for $\beta'$. If $\eta' \leq \lambda$, the simulator $\mathcal{B}'$ does not artificially abort.

We assume that the simulator makes $\mathcal{O}(\varepsilon^{-2} \ln(\varepsilon^{-1}) \lambda^{-1} \ln(\lambda^{-1}))$ samples. Using Chernoff's bound, one obtains the following bound for the estimate $\eta'$ of $\eta$ :
$$\Pr[|\eta' - \eta| > \eta \cdot \varepsilon/8] < \lambda \frac{\varepsilon}{8} \tag{4}$$

Let Art denote the event that the simulator artificially aborts in the guess phase. Let Abort denote the event that the simulator aborts, either artificially or not. Then for a fixed parameter $\mathfrak{P}$, we have :

$$\Pr[\overline{\mathsf{abort}}] = \Pr_{\boldsymbol{X},k}[\tau(\boldsymbol{X},\boldsymbol{v},v^*,k) = 0] \cdot \Pr[\overline{\mathsf{Art}}] = \eta \cdot \Pr[\overline{\mathsf{Art}}]$$

We distinguish two cases : if $\eta \cdot (1 - \varepsilon/8) > \lambda$, then for a fixed $\eta'$, if $|\eta' - \eta| \leq \eta \cdot \varepsilon/8$, we have that $\eta' > \lambda$ which gives $\Pr[\overline{\mathsf{Art}}] = \lambda/\eta'$ and :

$$\lambda \cdot (1 - \varepsilon/8) \leq \frac{\lambda}{1 + \varepsilon/8} \leq \Pr[\overline{\mathsf{Abort}}] \leq \frac{\lambda}{1 - \varepsilon/8} \leq \lambda \cdot (1 + \varepsilon/4)$$

Then for a randomly sampled $\eta'$, we obtain using (4) :

$$\left|\Pr[\overline{\mathsf{Abort}}] - \lambda\right| \leq \lambda \cdot \varepsilon/2 \tag{5}$$

One can show that the same inequality holds if $\eta \cdot (1 - \varepsilon/8) \leq \lambda$. Then since inequality (5) holds for any fixed parameter $\mathfrak{P}$, this remains true for a random $\mathfrak{P}$, conditioned on $\gamma' = \gamma$ or $\gamma' \neq \gamma$, that is :

$$\left|\Pr[\overline{\mathsf{abort}}|\gamma' = \gamma] - \lambda\right| \leq \lambda \cdot \varepsilon/2 \tag{6}$$
$$\left|\Pr[\overline{\mathsf{abort}}|\gamma' \neq \gamma] - \lambda\right| \leq \lambda \cdot \varepsilon/2 \tag{7}$$

We have that if $\beta = 1$, the simulator succeeds if it outputs $\beta' = 1$; this happens if it aborts and then correctly guesses $\beta'$ (with probability $\frac{1}{2}$), or if it does not abort and $\gamma = \gamma'$ :

$$\Pr[\beta' = 1] = \Pr[\beta' = 1 \wedge \mathsf{abort}] + \Pr[\beta' = 1 \wedge \overline{\mathsf{abort}}]$$
$$= \frac{1}{2} \cdot (1 - \Pr[\overline{\mathsf{abort}}]) + \Pr[\gamma' = \gamma \wedge \overline{\mathsf{abort}}]$$
$$= \frac{1}{2} + \frac{1}{2}\left(\Pr[\gamma' = \gamma \wedge \overline{\mathsf{abort}}] - \Pr[\gamma' \neq \gamma \wedge \overline{\mathsf{abort}}]\right)$$
$$= \frac{1}{2} + \frac{1}{2}\left(\Pr[\overline{\mathsf{abort}}|\gamma' = \gamma] \cdot \Pr[\gamma' = \gamma] - \Pr[\overline{\mathsf{abort}}|\gamma' \neq \gamma] \cdot \Pr[\gamma' \neq \gamma]\right)$$

Using inequalities (6) and (7), we obtain :

$$\left|\Pr[\beta' = 1|\beta = 1] - \frac{1}{2}\right| \geq \frac{\lambda}{2}\left|\Pr[\gamma' = \gamma] - \Pr[\gamma' \neq \gamma]\right| - \frac{\lambda}{2} \cdot \varepsilon$$

Since the adversary is a $(t, q, \varepsilon)$-adversary, we have that :

$$\left|\Pr[\gamma' = \gamma] - \frac{1}{2}\right| \geq \varepsilon$$

This gives :

$$\left|\Pr[\beta' = 1|\beta = 1] - \frac{1}{2}\right| \geq \lambda \left|\Pr[\gamma' = \gamma] - \frac{1}{2}\right| - \frac{\lambda}{2} \cdot \varepsilon \geq \frac{\lambda}{2} \cdot \varepsilon$$

Finally, when $\beta = 0$, the simulator either aborts and outputs a random $\beta'$, or does not abort and outputs $\beta' = 1$ if $\gamma = \gamma'$. Since for $\beta = 0$ the adversary has no information about $\gamma$, we obtain in both cases :

$$\Pr[\beta' = 1 | \beta = 0] = \frac{1}{2}$$

This gives :

$$\left| \Pr[\beta' = 1] - \frac{1}{2} \right| \geq \frac{1}{4} \cdot \lambda \cdot \varepsilon \geq \frac{\varepsilon}{16 \cdot q \cdot 2^\ell \cdot n}$$

which terminates the proof. □

## 6 All in All...

The previous theorem shows that the probability of breaking the new IBE scheme is lesser than $q \cdot 2^{\ell+4} \cdot n$ times the probability of solving the DBDH problem under similar timing constraints; recall that $\ell$ is the factor by which the public parameter size is divided compared to Waters' scheme. When $\ell = 1$ we recover the security level of Waters' scheme. This means that when the public parameter size is divided by a factor $\ell$, the security gets reduced by $\ell$ bits. Thus, one needs to strike a trade-off between security and public parameter size.

We recommend to take $\ell = 32$. This means that we loose 32 bits of security compared to Waters' scheme (this is so slight for all engineering purposes one can conisder that no security is lost). Note however that this does not necessarily mean that there exists an attack against the new scheme that is $2^{32}$ faster than the best attack against Waters' scheme. Currently the best known attack against both schemes consists in solving the discrete-logarithm problem in $\mathbb{G}$. When $\mathbb{G}$ is the group of points of a well chosen elliptic curve, this requires exponential time. Actually, those 32 security bits correspond to the difference between the security level that can be guaranteed for Waters' scheme and the security level that can be guaranteed for the new scheme; again, this does not necessarily mean that the new scheme is "less secure" than Waters' scheme in practice.

With $\ell = 32$ and $n' = n \cdot \ell = 160$, we obtain $n = 5$ (instead of $n = 160$ in Waters' scheme), and the public parameter size is still $n + 4$ group elements.

Therefore, when an elliptic curve over $\mathbb{F}_p^2$ is used (the simpler setting) with a 512-bit prime $p$, the public parameter size is nine kilobytes (4.5 kilobytes with compressed points) instead of 164 kilobytes (82 kilobytes with compressed points).

To further compensate security, one can also use a larger prime $p$. For example, with a 1024-bit prime $p$, the public parameter size becomes eighteen kilobytes (nine kilobytes with compressed points).

In conclusion, we have introduced a variant of Waters' identity-based encryption scheme with a much smaller public-key size (only a few kilobytes). We proved that this variant is a fully secure in the standard model. The new scheme allows to compress the public parameter size by a multiplicative factor of $\ell$, at the cost of reducing the security level by only $\ell$ bits.

Therefore, the new scheme constitutes the *first fully secure and fully practical identity-based encryption scheme known to date*.

## A  Bounding The Abortion Probability

We want to show that :

$$\Pr_{\boldsymbol{X},k}[\tau(\boldsymbol{X},\boldsymbol{v},v^*,k)=0] \geq \lambda = \frac{1}{4 \cdot q \cdot 2^\ell \cdot n}$$

where :

$$\tau(\boldsymbol{X},\boldsymbol{v},v^*,k) = \begin{cases} 0, & \text{if } F(v^*)=0 \text{ and } F(v^j) \neq 0 \mod m \text{ for all } 1 \leq j \leq q \\ 1, & \text{otherwise} \end{cases}$$

First, we define the modified function $\tau'(\boldsymbol{X},\boldsymbol{v},v^*)$ :

$$\tau'(\boldsymbol{X},\boldsymbol{v},v^*) = \begin{cases} 0, & \text{if } F(v^*)=0 \mod m \\ & \text{and} \\ & F(v^j) \neq 0 \mod m \text{ for all } 1 \leq j \leq q \\ 1, & \text{otherwise} \end{cases}$$

$\forall v$, we have that

$$0 \leq x' + \sum_{i=1}^{n} v_i \cdot x_i \leq m - 1 + n \cdot (2^\ell - 1) \cdot (m - 1) < m \cdot n \cdot 2^\ell$$

This shows that if $F(v^*) = 0 \mod m$, then there is a unique $0 \leq k < n \cdot 2^\ell$ such that $F(v^*) = 0$ over the integers. Since $k$ is uniformly and independently distributed between zero and $2^\ell \cdot n - 1$, we have that:

$$\Pr_{\boldsymbol{X},k}[\tau(\boldsymbol{X},\boldsymbol{v},v^*,k) = 0] = \frac{1}{2^\ell \cdot n} \cdot \Pr_{\boldsymbol{X}}[\tau'(\boldsymbol{X},\boldsymbol{v},v^*) = 0] \qquad (8)$$

We denote by $A_j$ the event that the simulator can answer the $j$-th private-key query:

$$\mathsf{A_j} : \sum_{i=1}^{n} v_i^j \cdot x_i \neq 0 \mod m$$

and by $\mathsf{B}'$ the event:

$$\mathsf{B}' : \sum_{i=1}^{n} v_i^* \cdot x_i = 0 \mod m$$

We have that:

$$\Pr_{\boldsymbol{X}}[\tau'(\boldsymbol{X},\boldsymbol{v},v^*) = 0] = \Pr[\bigwedge_{j=1}^{q} \mathsf{A_j} \wedge \mathsf{B}'] \qquad (9)$$

We have:

$$\Pr[\bigwedge_{j=1}^{q} \mathsf{A_j}|\mathsf{B}'] = 1 - \Pr[\bigvee_{j=1}^{q} \neg\mathsf{A_j}|\mathsf{B}'] \geq 1 - \sum_{j=1}^{q} \Pr[\neg\mathsf{A_j}|\mathsf{B}'] \qquad (10)$$

The pairwise independence of the function $F(v) \mod m$ stems from the following lemma, which proof is straightforward:

**Lemma 1.** *Let $x'$ and $x_1, \ldots, x_n$ be random variables uniformly distributed between zero and $m - 1$ and let $F(v) = x' + \sum_{i=1}^{n} v_i \cdot x_i$. Then for all $v \neq v'$ and all $a, a' \in \mathbb{Z}_m$,*

$$\Pr[(F(v) = a \mod m) \wedge (F(v') = a' \mod m)] = \frac{1}{m^2}$$

The previous lemma shows that $\Pr[\mathsf{B}'] = 1/m$ and that for all $j$:

$$\Pr[\neg\mathsf{A_j}|\mathsf{B}'] = \frac{\Pr[\neg\mathsf{A_j} \wedge \mathsf{B}']}{\Pr[\mathsf{B}']} = \frac{1}{m}$$

Then using (8), (9), (10), we obtain:

$$\Pr_{\boldsymbol{X},k}[\tau(\boldsymbol{X},\boldsymbol{v},v^*,k) = 0] \geq \frac{1}{2^\ell \cdot n} \cdot \left(1 - \frac{q}{m}\right) \frac{1}{m}$$

And $m = 2 \cdot q$ yields the following lower bound $\lambda$:

$$\Pr_{\boldsymbol{X},k}[\tau(\boldsymbol{X},\boldsymbol{v},v^*,k) = 0] \geq \lambda = \frac{1}{4 \cdot q \cdot 2^\ell \cdot n}$$